**Refined Characterization of Student Perspectives on Quantum Physics**

Charles Baily and Noah D. Finkelstein

Department of Physics, University of Colorado, Boulder, Colorado 80309, USA

**ABSTRACT**

The perspectives of introductory classical physics students can often negatively influence how those students later interpret quantum phenomena when taking an introductory course in modern physics. A detailed exploration of student perspectives on the interpretation of quantum physics is needed, both to characterize student understanding of physics concepts, and to inform how we might teach traditional content. Our previous investigations of student perspectives on quantum physics have indicated they can be highly nuanced, and may vary both within and across contexts. In order to better understand the contextual and often seemingly contradictory stances of students on matters of interpretation, we interviewed 19 students from four introductory modern physics courses taught at the University of Colorado. We find that students have attitudes and opinions that often parallel the stances of expert physicists when arguing for their favored interpretations of quantum mechanics, allowing for more nuanced characterizations of student perspectives in terms of three key interpretive themes. We present a framework for characterizing student perspectives on quantum mechanics, and demonstrate its utility in interpreting the sometimes-contradictory nature of student responses to previous surveys. We further find that students most often vacillate in their responses when what makes intuitive sense to them is not in agreement with what they consider to be a correct response, underscoring the need to distinguish between the *personal* and the *public* perspectives of introductory modern physics students.

# I. INTRODUCTION

Prior research has indicated that through instruction in classical physics, or from everyday experience, many introductory physics students develop realist perspectives [1] based in part on intuitive conceptions of particle and wave phenomena. For introductory modern physics students, we have seen how realist perspectives may translate into specific beliefs about quantum phenomena: e.g., quanta are always localized in space, or that probabilistic descriptions of quantum measurements are statistical and incomplete. It has been shown, however, that no local theory (including local realism) can reproduce all the predictions of quantum mechanics, [2] making questions of interpretation scientifically relevant, subject to experimental test, and not simply a matter of philosophical taste. A detailed exploration of student perspectives on quantum physics is therefore necessary, since these perspectives are an aspect of understanding physics, and have implications for how traditional content in quantum mechanics might be taught. Introductory modern physics courses are of particular interest since they often represent a first opportunity to transition students away from classical epistemologies and ontologies, to ones that are more aligned with the beliefs of contemporary physicists.

Still, it is not always clear exactly what expert physicists believe regarding the interpretation of quantum mechanics. [3] A recent survey [4] of quantum physics instructors at the University of Colorado and elsewhere (all of whom use quantum mechanics in their research) found that 30% of those surveyed interpreted the wave function as being physically real, while nearly half considered it to contain information only. The remaining respondents held some kind of mixed view on the physical interpretation of the wave function, or saw little distinction between the two choices. And only half of those who expressed a clear preference in the survey (matter-wave or information-wave) did so with confidence, and were of the opinion that the other view was probably wrong. Not surprisingly, modern physics instructors vary in whether and how to teach matters of interpretation to introductory quantum physics students, with a demonstrable impact on student thinking. For example, students have been found to be less likely to prefer realist interpretations of quantum systems when instructors are explicit in promoting alternative perspectives. [1,5]

Education research in quantum physics has primarily focused on student difficulties with the mathematical formalism of quantum mechanics, [6, 7] such as the time evolution of quantum states or the calculation of expectation values, [8] with relatively little attention paid to student understanding or instructional practices relevant to interpretive matters. [9, 10] Some research has explored the models employed by students in describing quantum phenomena, but have been generally confined to highly specific contexts, [11, 12] and framed more in terms of student sophistication with scientific modeling than student understanding of the interpretative aspects of quantum mechanics. Mannila et al. [13] have employed a complex interpretive framework in exploring student perspectives on quantum phenomena in the specific context of a single-quanta double-slit experiment. These researchers found that open-ended student responses to a series of written questions were dominated by "semi-classical" or "trajectory-based" ontologies, and that very few students expressed perspectives that were aligned with expert models, or even productive transitional models (nonlocal and/or statistical perspectives, by their standards); they also reported many instances of mixed student ontologies within that specific context of a single-quanta double-slit experiment. Our studies have indicated that student perspectives can be highly sensitive to context, so that it may not always be possible to make generalizations about student beliefs based on explorations within a single problem statement. We are currently unaware of education research addressing broad characterizations of the interpretive aspects of student perspectives on quantum mechanics, nor of any that adequately address specific reasons for the contextually based or often contradictory stances of students with respect to interpretive matters in quantum physics.

Our previous efforts to characterize student perspectives on quantum physics were limited to the application of three coarse labels (*Realist*, *Quantum*, or *Agnostic*, to be discussed below). These categorizations were based on an analysis of student responses to a post-instruction online essay question on interpretations of the double-slit experiment, coupled with responses to a survey statement concerning the existence of an electron's position within an atom. [See the Appendix and the discussion in Sec. III below.] This classification scheme is limited in terms of capturing the many nuances of student responses, and in understanding why students seem to exhibit contradictory perspectives between and within these two specific contexts. In this paper, we therefore address the following questions:

1) How might previous classification schemes be refined to better describe the nuances of student perspectives on interpretive themes in quantum physics?
2) For what reasons do students exhibit mixed perspectives within and across contexts?

From a total of 19 post-instruction interviews with students from four recent introductory modern physics courses at the University of Colorado we find that, though they may not employ the same formal language as expert physicists, students often invoke concepts and beliefs that parallel those invoked by expert physicists when arguing for their preferred interpretations of quantum mechanics. These parallels allow us to characterize student perspectives on quantum physics in terms of some of the same themes that distinguish these formal interpretations from each other. Of particular significance is the finding that students do indeed have attitudes and opinions regarding these various themes of interpretation, regardless of whether these themes had been explicitly addressed by their instructors in class. With regard to mixed or seemingly contradictory responses from students, results from the present studies indicate that: (i) Some students prefer a mixed wave-particle ontology (a pilot-wave interpretation, wherein quanta are simultaneously *both* particle *and* wave); and (ii) Students are most likely to vacillate in their responses when what makes intuitive sense to them is not in agreement with what they consider to be a correct response.

**II. INTERVIEW PARTICIPANTS AND COURSE CHARACTERISTICS**

Observations of varying instructional approaches regarding matters of interpretation in modern physics courses taught at the University of Colorado (and their associated impacts on student thinking**)** have led to our current efforts to more deeply explore student perspectives on quantum mechanics. To do so, we sought to recruit five students from each of four modern physics offerings from a single academic year to participate in an hour-long post-instruction interview. A mass email was sent to all students enrolled in these courses, offering fifteen dollars in exchange for their time; students were not informed ahead of time about the nature of the interview questions, only that we would be discussing some ideas from modern physics. There was no real opportunity to select among students since volunteers were sometimes scarce, and so there was no attempt to make the cohort representative of all students from those courses.

| INSTRUCTOR | STUDENT POPULATION | COURSE/INSTRUCTOR INTERPRETIVE STANCE | # OF STUDENTS INTERVIEWED |
|---|---|---|---|
| Red | Engineering | Matter-Wave | 3 |
| Blue | Engineering | Copenhagen | 5 |
| Green | Physics | Copenhagen/Agnostic | 6 |
| Yellow | Physics | Copenhagen/Agnostic | 5 |

TABLE I. Summary of the four courses from which students were recruited for interviews, including a characterization of each instructor's stance on interpretive themes as taught in that course. A total of nineteen students were interviewed.

A total of 19 students were interviewed from these four courses, as summarized in Table I, either in the last week of the semester or after the course had ended. Interview participants from the courses for physics majors were all physics or engineering physics majors, plus one astronomy major; those from the courses for engineers were all engineering majors (but not engineering physics), plus one mathematics major. The average final course grade for all 19 students was 3.4 (out of 4.0, where overall course averages fall in the 2.0–3.0 range), indicating that participants were generally better than average students, as might be expected for a group of volunteers. Neither of the authors of this paper were instructors for any of these courses, and none of the instructors were physics education researchers.

Our characterizations of the various interpretive stances of theses instructors, which come from analyses of course materials and practices, have been described in prior work, [5] and can be best illustrated by how an instructor addressed the double-slit experiment. We emphasize that our characterization of an interpretive stance is not necessarily a reflection of the instructor's personal perspective on quantum mechanics, but rather how that instructor addressed interpretative themes during class.

Instructor Red was explicit in promoting a wave model of individual electrons, telling students that each electron exists as a delocalized wave as it propagates through both slits, interferes with itself, and then becomes localized upon detection. Instructor Blue told students that a "quantum mechanical wave of probability" passes through both slits, but that which-path questions are unanswerable without disrupting the interference pattern. While similar to Instructor Blue, Instructors Green and Yellow ultimately placed more emphasis on calculation (predicting features of the interference pattern) than matters of interpretation.

Analyses of aggregate student responses to an end-of-term online survey described previously [5] illustrate both the differential impact of these instructional approaches on student thinking, as well as the mixed nature of student responses across contexts. The results may be summarized as follows:

1) Instructors were generally explicit in teaching a particular interpretation of the double-slit experiment with single quanta, *though not explicitly as an interpretation*; student responses in this context were overall reflective of the teaching goals for each course. Students from the Matter-Wave course overwhelmingly preferred a wave-packet (Quantum) description of electrons in the double-slit experiment (each electron passes through both slits and interferes with itself), while responses from the three Copenhagen/Agnostic courses were more varied. These latter students were not only more likely to prefer an Agnostic stance (quantum mechanics is about predicting the interference pattern, not discussing what happens in between), they were also more likely to align themselves with a Realist interpretation (each electron goes through either one slit or the other, but not both). Among the responses from all four courses were a small number (~5% of total students) who chose to agree with *both* Realist and Quantum interpretations of the double-slit experiment. From this we conclude that instructor practices do influence student thinking.

2) Instructors were considerably less explicit about interpretation at later stages of the course, as when students learned about the Schrödinger model of hydrogen, and a large fraction of students from all four courses chose to agree with the statement: *An electron in an atom has a definite but unknown position at each moment in time*. Disagreement with this statement could be consistent with either a Quantum or an Agnostic perspective, whereas agreement would be more consistent with a Realist perspective. Students often do not generalize their interpretive frameworks across contexts.

3) When student responses are combined so that responses to the statement on atomic electrons are grouped by how those same students responded to the essay question on the double-slit experiment, students in the Realist category were the most consistent, with most preferring realist interpretations in both contexts. However, nearly half of the students who preferred a wave-packet description of electrons in the double-slit experiment would still agree that electrons in atoms exist as localized particles.

For the 19 students interviewed for the present studies, there were no discernible connections between a particular instructional approach and the preferred perspectives of the students from that course, likely due to the limited number of participants. Therefore, discussion in this paper of specific instructional approaches will be limited to the brief characterizations given above, and a few specific statements below concerning the influence of a certain instructional approach on student responses.

## III. REFINING CHARACTERIZATIONS OF STUDENT PERSPECTIVES

With the objective of improving on our former coarse-grained classification scheme, we have found it useful to consider student perspectives in terms of concepts associated with some of the more common (i.e., less exotic) formal interpretations of quantum mechanics. In doing so, we do not mean to imply that student perspectives on quantum physics are as coherent or sophisticated as any formal interpretation (although other research [4] suggests that expert perspectives on quantum physics may be similarly tentative). In fact, a conceptions model [14] of student perspectives would be inadequate in accounting for the contextual nature of student responses, which would be more consistent with a resources model [15] of cognition and student learning. Nor do we assume that any one formal interpretation is necessarily sufficient for describing the nuanced and sometimes-inconsistent perspectives exhibited by any particular student, or even that the development of student perspectives on quantum physics follows along the lines of historical developments. We do, however, find that some formal interpretations of quantum mechanics can be distinguished from each other in terms of a few key themes (discussed below), and that students do have attitudes and ideas concerning these themes of interpretation, regardless of whether these themes had been explicitly addressed by their instructors. In other words, we have observed that many introductory modern physics students, when formulating a stance on these interpretive themes, employ some of the same epistemological tools used by expert physicists, and will sometimes invoke similar experimental results and intuitive notions of particles and waves as motivation for their preferred interpretations of quantum phenomena.

From our interviews, we find that students express beliefs and attitudes (of varying degrees of sophistication) concerning the following three interpretive themes, which emerged during an analysis of all interview transcripts:

1) Is the position of a particle objectively real, or indeterminate and observation dependent? (Existence or nonexistence of certain hidden variables.)
2) Is the wave function a mathematical tool that encodes probabilities (information-wave), or is it physically real (matter-wave)?
3) Does the collapse of the wave function (or reduction of the state) represent a physical process, or simply a change in knowledge of the observer?

### III.A. Discussion of Formal Interpretations

In Table II we summarize our characterizations of several formal interpretations of quantum mechanics with respect to the three interpretive themes listed above; we then offer brief working descriptions of each interpretation below. We emphasize that it would be impossible for these descriptions to be comprehensive, and we anticipate that some may disagree with certain aspects of our characterizations, but we offer these working definitions for the sake of clarity when associating these labels with the expressed beliefs of specific students.

| INTERPRETATION | HIDDEN VARIABLES? | INFO OR MATTER WAVE? | COLLAPSING WAVE FUNCTION? |
|---|---|---|---|
| REALIST/STATISTICAL | YES/AGNOSTIC | INFO | KNOWLEDGE |
| COPENHAGEN | NO | INFO | PHYSICAL |
| MATTER-WAVE | NO | MATTER | PHYSICAL |
| PILOT-WAVE | YES | MATTER | KNOWLEDGE |
| AGNOSTIC | AGNOSTIC | AGNOSTIC | AGNOSTIC |

TABLE II. A summary of our characterizations of four formal interpretations of quantum mechanics, in terms of three interpretive themes. The Agnostic perspective is not considered to be a formal interpretation in itself, but is included for completeness.

**Realist/Statistical** – From either a Realist or Statistical perspective, the physical properties of a system are objectively real and independent of experimental observation. The state vector encodes probabilities for the outcomes of measurements performed on an ensemble of similarly prepared systems, but cannot provide a complete description of individual systems. The wave function is not physically real; the *collapse of the wave function* represents a change in the observer's knowledge of the system, and not a physical change brought about by the act of measurement.

**Copenhagen** – The probabilistic nature of quantum measurements is a reflection of the inherently probabilistic behavior of quantum entities; in general, the properties of a system are indeterminate until measured. The wave function is not a literal representation of a physical system, but the collapse of the wave function does represent a physical transition from an indeterminate state to one where certain properties of the state become well defined. Speculations about physical processes that are unobservable would be considered outside the domain of science.

**Matter-Wave** – Similar to the Copenhagen Interpretation with respect to indeterminacy and the nonexistence of hidden variables, but also ascribes physical reality to the wave function. Though not described by the Schrödinger equation, the collapse of the wave function represents a physical process induced by measurement. [Student responses consistent with a matter-wave interpretation are characterized in this paper as *Quantum* to be consistent with prior studies.]

**Pilot-Wave** – From this perspective, quanta are simultaneously both particle and wave: localized particles follow trajectories determined by a physically real wave function. In the double-slit experiment, an electron is all at once both a particle that goes through only one slit, and a wave that passes through both slits and interferes with itself. In this context, the position of a particle is objectively real and predetermined based on unknowable initial conditions, so that the reduction of the state represents a change in knowledge of the observer.

**Agnostic** – Though not a formal interpretation in itself, we distinguish between this stance and the positivistic aspects of the Copenhagen Interpretation (declining to speculate on the unobservable). The Agnostic perspective takes into account multiple interpretations of quantum mechanics and their ontological implications, but takes no definite stance on which might correspond to the best description of reality. The utility of quantum mechanics is generally favored over its interpretive aspects.

**III.B. Students Express Beliefs that Parallel Those of Expert Proponents**

The perspectives of many modern physics students on quantum phenomena are significantly influenced by the commonplace notion of particles as localized in space. In classical physics, as in colloquial usage, the word *particle* generally connotes some small object, so it should not be surprising that students who have learned about particles primarily within the context of classical physics should persist in thinking of them as microscopic analogs to macroscopic objects when learning about quantum physics. In doing so, students may consider the position of an electron in an atom to be as much a quality of that particle as its mass or charge [Student codes are as given in Table III]:

> "I guess an electron has to [always be at] a definite point. It is a particle, we've found it has mass and it has these intrinsic qualities, like the charge it has, so it will have a definite position, but due to uncertainty it will be a position that is unknown." [STUDENT QR2]

This statement reveals not only one student's belief in localized massive particles, it also suggests a stance on the uncertainty associated with a particle's position – its objectively real value will be unknown until revealed by measurement. This student (and others with similar attitudes) interpreted the probability density for an atomic electron as strictly a mathematical tool describing only the probable locations for where that electron might be found once measured; probabilistic descriptions of such measurements were therefore seen as a reflection of ignorance concerning the true state of that particle just prior to measurement. We thus see how an intuitive notion of particles as localized objects can influence what physical meaning students ascribe to both the wave function and the probabilistic nature of quantum mechanics.

In a similar vein, another student objected to the idea that wave packets could represent single particles. Here, this student is discussing the Quantum Wave Interference (QWI) simulation, [16] which depicts the propagation of a wave packet through both of two slits on its way to the detection screen:

> "One electron can't go through both slits at the same time because electrons have mass. Wouldn't it violate conservation of mass and charge if [the electron] were split into two like it shows in the [QWI] simulation?" [STUDENT R1]

Such objections are reminiscent of Ballentine (a major proponent of the Statistical Interpretation of quantum mechanics [17]) when discussing a thought experiment, in which an incident wave packet is divided by a semi-reflecting barrier into two distinct transmitted and reflected wave packets. The reflected and transmitted waves are then directed toward a pair of detectors connected to a coincidence counter. Ballentine argues:

> "Suppose that the wave packet *is* the particle. Then since each packet is divided in half, […] the two detectors will always be simultaneously triggered by the two portions of the divided wave packet." [17, p.101, emphasis in original.]

In this thought experiment (and in practice [18]), individual quanta trigger either one detector or the other (and not both simultaneously); Ballentine therefore concludes that, while the wave function may have nonzero amplitude in two spatially separated regions, it cannot be interpreted as describing individual particles since individual particles are never found in two places at once. In making this argument, Ballentine implicitly assumes that the *collapse of the wave function* (or *reduction of the state*) represents a change in knowledge of the observer, and not an actual physical process induced by measurement.

In his own book on quantum mechanics, Dirac [19] considers the same type of thought experiment as Ballentine, but provides a radically different explanation:

> "The result of [the detection] must be either the whole photon or nothing at all. Thus the photon must change suddenly from being partly in one beam and partly in the other to being entirely in one of the beams." [19, p. 9]

As unintuitive as this interpretation may be, we find that a number of modern physics students report having accepted such ideas, and have incorporated them into their descriptions of quanta:

> "[T]he electron, until it's measured, until you try to figure out where it is, the electron is playing out all the possibilities of where it could go. Once you measure where it is, that collapses its wave function [and it] loses its properties as a wave and becomes particle in nature." [STUDENT Q3]

Whereas other students, like Ballentine, find these types of explanations unsatisfying:

> "[A] single electron is detected at the far screen, and I feel like that really can't be explained for the wave packet, by one specific detection in a small place like that, if you say [the wave packet] is the electron. That's really the only discrepancy I have with that: What happens when it hits the screen?" [STUDENT QR2]

Indeed, the question of what happens when individual quanta are detected in the double-slit experiment has played a significant role for some physicists in motivating their perspectives on quantum phenomena, as with Ballentine:

> "[I]t is possible to detect the arrival of individual electrons, and to see the diffraction pattern emerge as a statistical pattern made up of many small spots. *Evidently, quantum particles are indeed particles*, but particles whose behavior is very different from what classical physics would have led us to expect." [17, p. 4, emphasis added]

J. S. Bell has also invoked the double-slit experiment when discussing interpretation, but in this case as motivation for a pilot-wave interpretation, as proposed by Bohm and others [20]:

> "Is it not clear from the smallness of the scintillation on the screen that we have to do with a particle? And is it not clear, from the diffraction and interference patterns, that the motion of the particle is directed by a wave?" [2, p. 191]

This student's discussion of the double-slit experiment echoes sentiments of both Bell and Ballentine:

> "[F]or me, saying that the [wave] represents the electron isn't accurate because an electron, after it's measured on that screen, is a point-particle, you see a distinct interference pattern after shooting many electrons, but you still see one electron hit the screen individually. […] I do agree that the electron acts as a wave because that's obviously what causes the pattern; if it didn't interfere with itself, or create a wavelike function, then you wouldn't see the patterns on the screen also." [STUDENT R3]

Historically, and in our classrooms today, different physicists have offered different interpretations of quantum diffraction experiments. For Ballentine, diffraction patterns form as a consequence of the quantized momentum transfer between localized particles and the diffracting medium. [17, p. 136] These patterns are more commonly explained in terms of wave interference, but for some the wave is guiding the trajectory of a localized particle, while others would claim that each particle interferes with itself as a delocalized wave until becoming localized upon detection. At the same time, a number of both expert and student physicists find it unscientific to speculate on that which cannot be experimentally observed:

> "I understand why people would think [the electron] has to exist between here and where it impacts, and it does, but the necessity of [thinking of it] between here and where it impacts as an actual concept like a particle or a wave, I don't see much of the point. We're not going to observe what it is between here and there, so it doesn't seem like a statement for science to make. It seems right now to be entirely unobservable." [STUDENT C2]

The refusal to speculate on unobservable processes is a key feature of the orthodox Copenhagen Interpretation of quantum mechanics, which seems to be favored by a majority of practicing physicists, if only for the fact that it allows them to apply the mathematical tools of the theory without having to worry about what's "really" going on (as embodied in the popular phrase: *Shut Up and Calculate!* [21]). Instructor Green communicated such agnostic sentiments to his students when faced with the question of whether an electron always has a definite but unknown position, or has no definite position until measured:

> "Newton's laws presume that particles have a well-defined position and momentum at all times. Einstein said that we can't know the position. Bohr said, philosophically, it has no position. *Most physicists today say: We don't go there. I don't care as long as I can calculate what I need.*" [INSTRUCTOR GREEN, emphasis added]

We also find it necessary to distinguish between the agnostic or positivistic aspects of an instructional approach, and the agnosticism of those who are aware of multiple interpretations, but are unsure as to which offers the best description of reality:

> "For now, for me, the electron is the wave function. But whether the electron is distributed among the wave function, and when you do an experiment, it sucks into one point, or whether it is indeed one particle at a point, statistically the average, I don't know." [STUDENT QA1]

**III.C. Categorization and Summary of Student Responses**

We summarize in Table III a categorization of individual students in terms of the interpretive themes discussed above, grouped by overall perspective, as discussed in Sec. III.A. A discussion of key findings and commonalities among students within individual categories follows. Interviews followed the protocol as given in the online supplementary materials.

We first note that many student responses agreed well with our characterizations of the formal interpretations, while other students provided one or more responses that were not entirely consistent with those characterizations; in some cases, a category was assigned based on what would be most consistent with the overall responses from that student. A second, independent physics education researcher coded a subset of five transcribed interviews (all students who were not quoted in this paper), both by interpretive theme and by overall interpretation, with an initial inter-rater reliability of 93% on individual stances on the interpretive themes, and 100% on overall perspective; following discussion, there was 100% agreement between both coders. All of the students in the *Split* category were explicit in distinguishing between what made intuitive sense to them (a Realist perspective) and what they perceived as a correct response (a Quantum perspective). Other students offered opinions on specific themes when asked to take a stance, but chose to ultimately remain agnostic for lack of sufficient information (as indicated by the XX/Agnostic entries in the interpretive themes columns of Table III). This characterization of individual responses differs from that of the general stance of Student QA1, who preferred a Quantum interpretation, but expressed a sophisticated overall agnosticism on the legitimacy of a contrasting Statistical Interpretation.

| STUDENT PERSPECTIVE | CODE | HIDDEN VARIABLES? | INFO OR MATTER WAVE? | COLLAPSING WAVE FUNCTION? |
|---|---|---|---|---|
| REALIST | R1 | YES | INFO | KNOWLEDGE |
|  | R2 | YES | INFO | KNOWLEDGE |
|  | R3 | YES | INFO | KNOWLEDGE |
| SPLIT QUANTUM/REALIST | QR1 | NO/YES | MATTER/INFO | PHYSICAL |
|  | QR2 | NO/YES | MATTER/INFO | KNOWLEDGE |
|  | QR3 | NO/YES | MATTER/INFO | KNOWLEDGE |
|  | QR4 | NO/YES | MATTER/INFO | AGNOSTIC |
| PILOT-WAVE | P1 | YES | MATTER | KNOWLEDGE |
|  | P2 | YES/AGNOSTIC | MATTER/AGNOSTIC | KNOWLEDGE/AGNOSTIC |
|  | P3 | YES | MATTER | KNOWLEDGE |
| QUANTUM | Q1 | NO | MATTER | KNOWLEDGE |
|  | Q2 | NO | MATTER | PHYSICAL/AGNOSTIC |
|  | Q3 | NO | MATTER | PHYSICAL |
|  | Q4 | NO | MATTER | PHYSICAL |
|  | Q5 | NO | MATTER | PHYSICAL/AGNOSTIC |
| QUANTUM/AGNOSTIC | QA1 | NO/AGNOSTIC | MATTER/AGNOSTIC | PHYSICAL/AGNOSTIC |
| COPENHAGEN | C1 | NO | INFO | PHYSICAL |
|  | C2 | NO | INFO/AGNOSTIC | PHYSICAL/AGNOSTIC |
|  | C3 | NO/AGNOSTIC | INFO | AGNOSTIC |

Table III. Summary of individual student interview responses with respect to three interpretive themes, grouped by overall perspective.

**III.C.1. Realist Perspectives**

All three of these students considered probability waves to be mathematical tools used only to describe the probable outcomes of measurements. These students all objected to the idea that a wave packet could represent a single particle, and said they always consider an electron to be a localized object traveling somewhere inside the probability wave describing the system. These students were not classified as holding a Statistical perspective because they were explicit in their belief in electrons as localized particles, and did not have sufficient content knowledge (e.g., consequences of Bell's Theorem) to appreciate why an agnostic stance on hidden variables might be desirable. All three of these students specifically objected to the notion of wave function collapse, calling it too counterintuitive or too unphysical to be a correct description of reality. These students were all aware of at least one alternative to their Realist interpretations, but said they hadn't yet been convinced by instructor arguments that their preferred perspective was incorrect.

**III.C.2. Split Quantum/Realist Perspectives**

While the Realist students above all expressed some confidence in their perspectives on quantum physics, even when those perspectives differed from what they had heard in class, the four students in this *split* Quantum-Realist category were, by the end of the interview, explicit in differentiating between what made intuitive sense to them, and what they considered to be a correct response. For example, Student QR1 first agreed that an electron in an atom always exists at a definite point, and continued with this line of thinking, both when first describing the double-slit experiment, and again as he began reading the Realist statement of (fictional) *Student One* from the double-slit essay question [See the interview protocol in Appendix A]:

> **STUDENT QR1:** I would agree with what Student One is saying, that the electron is traveling somewhere inside that probability density blob, and it is a tiny particle. The problem here that I see is that the electron went through one slit or the other. [PAUSE] So, now I'm disagreeing with myself. OK, my intuition is fighting me right now. I said earlier that there should be one point in here that is the electron, and it goes through here and hits the screen, but I also know that I've been told that the electron goes through both slits and that's what gives you the interference pattern. Interesting. [LONG PAUSE] OK, somehow I feel like the answer is going to be that this probability density, it is the electron, and that can go through both slits, and then when it's observed with this screen, the probability density wave collapses, and then only exists at one point. But at the same time I feel that there should be a single particle, and that somehow a single, finite particle exists in this wave, and will either travel through one slit or the other. Why would a single particle be affected by a slit? That I don't have an answer to, other than that it's the wave that's actually being propagated, the wave is the electron.

**INTERVIEWER:** OK. It seems like you're talking about two different ideas. One is that the electron is a point somewhere inside this wave, and the other is saying the electron is the wave. Do you feel those two ideas conflict in any way?

**QR1:** Yeah, they do, because one says there is a finite particle at all times, and the other says that there's not, there is just this probability density, and I think the answer will turn out to be that the electron is the probability density, and that's contrary to what I said earlier. But I don't see how it could be the other way, with a finite particle. I don't see how you could get an interference pattern here with the electron being a finite particle the whole time.

**INT:** OK. What about [the next statement]?

**QR1:** [BEGINS READING SECOND STATEMENT (a Quantum perspective)] So, that goes off of what I was just saying. [READS] So, I agree with everything up to here, the electron acts as a wave and will go through both slits and interfere with itself, I believe that's true. And that's why an interference pattern develops after shooting many electrons, I guess I agree with that too, because when the blob gets to the screen, it can't just still have a probability density that would look like an interference pattern by itself. It's going to have one finite location. But after multiple electrons, multiple blobs have passed through, they will collectively form an interference pattern. So I would agree with Student Two.

**INT:** So you're agreeing with Student Two. And did you say that you disagree with Student One, or do you just have reservations about what they're saying?

**QR1:** Intuitively, I kind of agree with Student One, but I think I have reservations. I don't think, Student One, that they're right.

**INT:** But it appeals to you, what they're saying?

**QR1:** Based upon lecture, and upon those who have greater knowledge of physics than I, I would say that this [second] statement agrees more with that than the initial situation.

**INT:** So you say Student One's statement disagrees more with what you've heard in class?

**QR1:** Yes. But not more with what I envision. This [first] one kind of depicts more of my rational depiction, all that I can wrap myself around and understand, and the second one is more of what I've been told, but don't completely understand. I've been told it's right, so…

This excerpt underscores the need to distinguish between the *personal* and the *public* perspectives [22] of students on quantum physics: these students differed from their Realist counterparts in that they explicitly differentiated between what made intuitive sense to them, and responses they perceived to be correct. This finding parallels studies by McCaskey, *et al.* [23] where students were asked to respond twice to the Force Concept Inventory. [25] first as they personally believed, and then as they felt a scientist would respond. These authors found that most every student *split* on at least one survey item, indicating a difference between their personal beliefs and their perceptions of scientists' beliefs. Following a series of validation interviews, these authors reported that students most often explained their personal responses in terms of what made intuitive sense to them, and that split responses reflected how students had learned a correct response from instruction, without having reconciled that knowledge with their own intuition. Similar studies probing the attitudes and beliefs of introductory classical physics students have demonstrated similar results. [26]

With respect to the public perspectives of modern physics students, we would also emphasize that students will not necessarily identify an authoritative stance based on specific knowledge of what expert physicists believe. Not only may their perceptions of what scientists believe be inaccurate, students may also employ undesirable epistemological strategies learned from their experiences in the classroom:

"This [Quantum statement of Student Two] is more of a complex definition, I think. […] Probably initially I would be confused by this statement if I hadn't taken this course, but I might be like the public and think the most complicated answer, that must be the right one. Because a lot of times – it's even happened with the [concept] questions in class – where I think: That's got to be the answer. But then I'll be like: No, that would be too easy, it's got to be something else. Sometimes that [strategy] can prove correct or incorrect…" [STUDENT QR4]

### III.C.3. Pilot-Wave Perspectives

The responses from these three students indicated a mixed wave-particle ontology. For these students, wave-particle duality meant that quanta must be thought of as simultaneously *both* wave *and* particle. The following student explained the fringe pattern in the double-slit experiment in terms of constructive and destructive interference, and acknowledged that the experiment had been used in class to demonstrate the wave characteristics of quanta, but had his own ideas about the source of interference for localized particles:

> "It seems like the probable paths for the electron to follow interact with themselves, but the electron itself follows just one of those paths. It's like the electron rides on a track, like a train rides on a rail, but those rails or tracks go through both slits, and the possible paths for the electrons to follow interfere with themselves, create the interference pattern, but the physical electron just rides on the tracks, it picks one. Or maybe switches paths, if two of them cross. I don't know, it seems that the electron has to be on one of those tracks, but the tracks themselves cause the interference pattern." [STUDENT P3]

Of particular interest is the way in which this same student demonstrated how his realist (albeit nonlocal) perspective can be used as an epistemological tool:

> "As [the electron is] traveling it's going to be somewhere in this [probability density] as it moves along until it's actually detected. And if it was here [INDICATES POINT NEAR DETECTING SCREEN], then it must have been here at one point in time [INDICATES SECOND POINT NEAR THE FIRST], and if it was here, then it had to be here at one point in time, all the way back to here [TRACES LINE BACK TO NEAR BOTH SLITS], in which case there's only two places it could be. So yes, I think it went through one slit or the other." [STUDENT P3]

As another example of how students offered inconsistent responses during interviews, one of these students explained that, while it is necessary to think of an electron in the double-slit experiment as both wave and particle, it was unnecessary to employ a wave description for atomic electrons since, in his mind, there were no wave effects to be accounted for:

> "When I was thinking about [an electron] in an atom, there's really no reason that you have to think about it as a wave, in the fact that it's not really interacting with anything. In [the double-slit] experiment, yes I like to think of it as also a wave, because this is kind of the key experiment of quantum mechanics, to describe this [wave] phenomenon, and so for that reason it is more effective to think of it as both." [STUDENT P1]

With this excerpt, we call attention to the fact that sometimes students employ different models for different contexts, without necessarily looking for internal consistency between them.

### III.C.4. Quantum Perspectives

These five students were consistent in providing responses that indicated a matter-wave ontology:

> "I don't think of [the electron] as orbiting the nucleus because it doesn't, it just exists in that region of space. It exists in a volume element that defines the probability of finding the electron in that space […] and that's really what the electron is: a smeared out volume of charge." [STUDENT Q2]

All of these students described unobserved quanta strictly in terms of waves, and discussed the collapse of the wave function as a physical process where wave-like quanta suddenly exhibit particle-like properties. In the minds of these students, their personal perspectives on quantum mechanics were in complete agreement with their perceptions of expert beliefs.

### III.C.5. Quantum/Agnostic Perspective

We distinguish this student from those in the strictly Quantum category, because while they had all expressed confidence in their matter-wave interpretations, this student expressed a degree of sophisticated uncertainty in his own views:

> "The way I think of an electron, I cannot ascribe to it any definite position, definite but unknown position. I mean, it may be that way, but I think that somehow the electron is represented by the wave function, which is just a probability, and if we want to localize it then we lose some of the information. So whether this is true or not is something of a philosophical question. I wish I knew, or understood it, but I don't. For now, for me, the electron is the wave function, so whether the electron is distributed among the wave function, and when you do an experiment, it sucks into one point, or whether it is indeed one particle at a point, statistically the average, I don't know." [STUDENT QA1]

### III.C.6. Copenhagen Perspectives

These three students were similar to the Quantum students in terms of the nonexistence of hidden variables, but saw probability waves as containing information only, rather than representing the actual physical state of a particle.  As with student C2 (quoted previously in Section III.B), each of these students stated explicitly that it is unscientific to discuss that which can't be measured or observed.  These three students considered electrons to be neither wave nor particle, stating that such concepts were in fact different models for describing the behavior of quanta under different circumstances.  One student explained how his own personal solipsistic philosophy influenced his beliefs about quantum mechanics, and vice-versa:

> **STUDENT C3:** This is more of a philosophical point for me, but if we can't know something, there's no difference between it not existing and us not knowing it.  So, for our purposes, it's more useful to say, if we can't know it, where the electron is, then it doesn't have a definite position. […]  I believe, so long as we don't measure it, then an electron doesn't have a definite position.
>
> **INTERVIEWER:** What happens when we measure it?
>
> **C3:** Well, we find a position then… Then it does.
>
> **INT:** The position we find, is that where the particle was the moment before we measured it?
>
> **C3:** No.  We can't know that.  So, when we make a measurement, there's the particle.  When we look away, the particle goes away.  And I sort of felt this way before having learned about quantum mechanics.  And it just solidified in my mind that there's no difference between me not knowing it, and it not existing.

With this excerpt, we would emphasize that formal instruction is not the only source of information or influence for students regarding quantum physics.  In the class-wide online surveys, a majority of students from all of the four courses considered here reported having previously heard about quantum mechanics in popular venues, such as books by Hawking [27] and Greene [28], before enrolling in the course.

## IV. DISCUSSION/CONCLUSIONS

We have improved upon our previous efforts to characterize the perspectives of introductory modern physics students on the nature of quantum physics through an exploration of their conceptions across three key interpretive themes. We find that students, as a form of sense-making, do develop ideas and opinions regarding the physical interpretation of quantum mechanics, regardless of whether and how their instructors explicitly address matters of interpretation in class. In exploring student perspectives on quantum physics, we find it natural that students would have attitudes regarding some interpretive themes, in that we were ultimately probing each student's ideas about the very nature of reality, and the role of science in describing it (questions of ontology and epistemology). Is the universe deterministic or inherently probabilistic? When is a particle a particle, and when is it a wave, and what is the nature of this wave? Is it unscientific to discuss the unobservable? Matters of interpretation are of both personal and academic interest to students, and modern physics instructors should recognize the potential impact on student thinking when choosing to de-emphasize interpretation in an introductory course. Not only do students develop their own ideas regarding the nature of quantum mechanics, they also develop attitudes about the positivistic or agnostic stances of their instructors:

> "It seems that there's this dogma among physicists, that you can't ask that question: What is it doing between point A and point B? 'You can't ask that!' And I think that the only way we'll be able to make profound progress is by asking those questions. It doesn't make sense that somebody would say, don't ask that, or you can't ask that. I think somehow they're shutting down free seeking of knowledge. But I don't know enough about quantum mechanics. Maybe when I get more understanding of quantum mechanics, I too will be saying: You can't ask that! But as a naïve student it sounds like a bad attitude to have about physics." [STUDENT P3]

Instructors may also want to recognize that not only are some questions of interpretation no longer a matter of philosophical taste (as evidenced by the work of Bell and others, [2, 29] but are also relevant to current modes of research in quantum information theory and elsewhere. Moreover, a common lament among physics education researchers is that we are losing physics majors in the first years of their studies by only teaching them 19th-century physics in our introductory courses. Similar issues may arise when modern physics instructors limit course content primarily to the state of knowledge in the first half of the last century. Although many instructors may argue that introductory students do not have the requisite sophistication to appreciate matters of interpretation in quantum mechanics, we note that several authors have developed discussions of EPR correlations and Bell inequalities that are appropriate for the introductory level. [30, 31] Questions of interpretation may also be addressed in terms of scientific modeling, an aspect of epistemological sophistication that is often emphasized in physics education research as a goal of instruction, as well as *nature of science* issues. In the end, we argue that modern physics instructors should concern themselves with matters of interpretation, if only because their students concern themselves with these matters, and as educators we should be concerned with what our students believe about physics and the nature of practicing physics. Modern physics instructors who aim to transition students away from classical epistemologies and ontologies may employ our framework for understanding and interpreting the myriad combinations of student ideas concerning the nature of the physical world, and our efforts as scientists to describe it.

As with other studies, we find that a significant number of students from our interviews (10 of 19) demonstrated a preference for realistic interpretations of quantum phenomena. However, the nature of these students' realist perspectives were not necessarily of the character we had anticipated from the results of earlier studies. Only three of the students who preferred local, realist interpretations of quantum phenomena expressed confidence in the correctness of their perspectives, whereas four others differentiated between what made intuitive sense to them (a Realist perspective) and what they perceived to be correct responses (a Quantum perspective). The inconsistent responses of these students may be understood in terms of their competing *personal* and *public* perspectives [22] on quantum physics – when responding in interviews or surveys, these students frequently vacillated between what they personally believed and the answer they felt an expert physicist would give, without always articulating a difference between the two without prompting. This finding has implications for future research into the ontologies of quantum physics students, who may not always respond to such questions as they actually believe, but rather provide responses that simply mimic their instructors. Of equal importance is the demonstration that these students may employ a variety of strategies in attempting to identify an expert response, some productive and some not. These issues are of particular significance with regard to matters of interpretation in quantum mechanics, where the beliefs of practicing physicists are at such variance with each other, which may confuse student perceptions.

The realist beliefs of three other students were of a decidedly nonlocal character: localized quantum entities follow trajectories determined by the interaction of nonlocal quantum waves with the environment. None of these three students claimed to be aware of any formal pilot-wave interpretation, and their beliefs in quanta as simultaneously wave and particle were at odds with how wave-particle duality was addressed in class by their instructors (i.e., quanta are sometimes described by waves, and sometimes as particles, but never both simultaneously).

The remaining nine of 19 students seemed to express fairly consistent views that could be viewed as in agreement with the instructional goals of their instructors – whether Quantum or Copenhagen. In other words, these students seemed to have successfully incorporated probabilistic and nonlocal views of quanta and quantum measurements into their personal perspectives, and/or agreed that scientists should restrict discussions to that which can be measured and verified. While these findings are somewhat at odds with previous research into quantum ontologies, which have found that student perspectives are rarely in alignment with expert or productive transitional models, we emphasize that the relatively few students who participated in our interviews were generally better-than-average students, and were not representative of an entire class. Ultimately, we believe the value of our findings lies in the demonstration and documentation of a variety of student beliefs regarding quantum phenomena, and not a determination of the prevalence of specific beliefs.

We also find it significant that most every student expressed distaste for deterministic ideas in the context of quantum physics, although it had been anticipated that Realist or Statistical students might favor such notions. Not only did most students say they were unfamiliar with the word *determinism* within the context of physics, practically every student believed either that the behavior of quantum particles is inherently probabilistic, or that the Uncertainty Principle places a fundamental limit on human knowledge of quantum systems, or a combination of both stances. We found that the Realist and the split Quantum-Realist students were more likely than other students to invoke the Uncertainty Principle when discussing notions of determinism; the remaining students were more likely to state that the behavior of quantum particles (or the nature of the universe**)** is inherently probabilistic. These responses indicate the need for a more detailed exploration of the Uncertainty Principle as an epistemological tool for students.

Our hope with this study is that an exploration of student perspectives on quantum physics will provide insight for both researchers and instructors into how students may be thinking about the content of modern physics courses, and the potentially negative impact of de-emphasizing interpretation in these courses. Such insight may allow us to target instructional interventions that will address student perspectives, thereby strengthening student abilities to make interpretations of physical phenomena, and to understand the limitations and bounds of these interpretations. A development of such interventions is the subject of current and future studies.

## V. ACKNOWLEDGEMENTS

The authors wish to thank the University of Colorado physics faculty members and students who helped facilitate this research, as well as Steve Goldhaber and the other members of the Physics Education Research group at Colorado for their helpful insights and suggestions. This work was supported in part by NSF CAREER Grant No. 0448176 and the University of Colorado.

# SUPPLEMENTARY MATERIALS

# Refined Characterization of Student Perspectives on Quantum Physics

## Charles Baily and Noah D. Finkelstein

**APPENDIX A – INTERVIEW PROTOCOL, NARRATIVE AND DISCUSSION**

**A.1. INTERVIEW PROTOCOL**

BACKGROUND INFORMATION
Name, declared major, previous physics and mathematics courses, both at CU and in high school, motivations for enrolling in the course

ASK STUDENTS TO DESCRIBE AN ELECTRON

DESCRIBE AN ELECTRON IN AN ATOM
Do students use a planetary model as a first-pass description? Are students aware of the limitations of the Bohr model? Do electrons move in orbits as localized particles? Does the student describe the electron in terms of an *electron cloud*, or a *cloud of probability*? What is this cloud? Does it represent something physical, or is it a mathematical tool? If the electron is described as a wave, what is it that's waving? Is there something moving up and down in space?

RESPOND TO THE STATEMENT:
*An electron in an atom has a definite but unknown position at each moment in time.*
IN AGREEMENT OR DISAGREEMENT AND EXPLAIN REASONING
Is the student's response consistent with their earlier descriptions of atomic electrons?

DESCRIBE THE SETUP FOR THE DOUBLE-SLIT EXPERIMENT
What is observed? Can the experiment be run with both light and electrons? What is observed when only single quanta pass through the apparatus at a time? What happens if you block one of the slits? What happens if you place a detector at one of the slits to see which slit individual quanta passed through? How do students explain the fringe pattern? If they explain the experiment in terms of localized particles, what is the source of interference? If they prefer a wave-description of quanta, how did the student explain why single quanta are detected as localized points? Is it possible for a particle to pass through two slits simultaneously? Does a wave packet description of individual particles reflect an ignorance of that particle's true position or momentum?

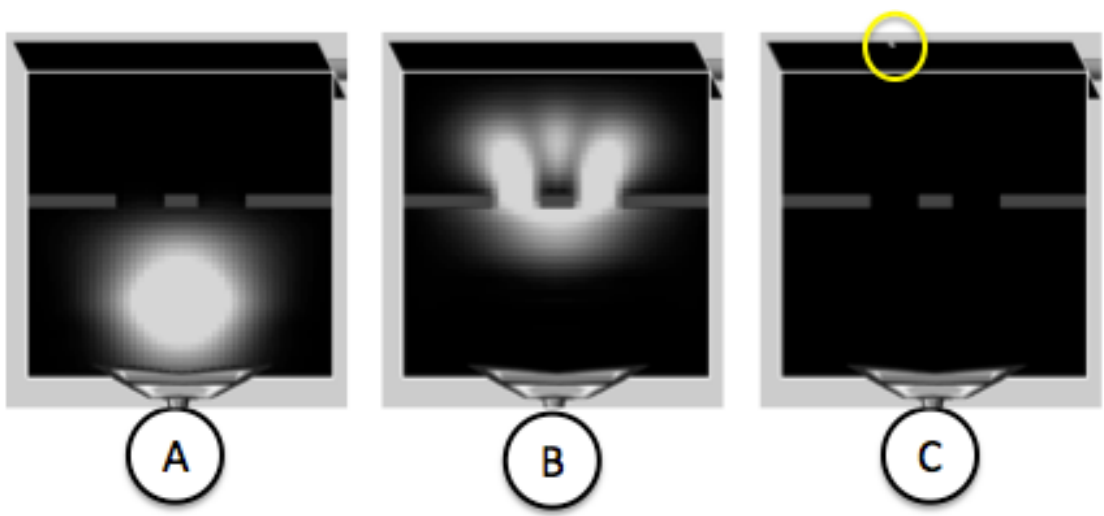

Fig. A1. A sequence of screen shots from the Quantum Wave Interference simulation. A bright spot (representing the probability density for a single electron) emerges from an electron gun (A), passes through both slits (B), and a single electron is detected on the far screen (C). After many electrons, a fringe pattern develops (not shown).

RESPOND TO THE ONLINE SURVEY QUESTION ON THE DOUBLE-SLIT EXPERIMENT:

> **Three students discuss the Quantum Wave Interference simulation (as depicted in Fig. A1):**
>
> **Student 1**: The probability density is so large because we don't know the true position of the electron. Since only a single dot at a time appears on the detecting screen, the electron must have been a tiny particle, traveling somewhere inside that blob, so that the electron went through one slit or the other on its way to the point where it was detected.
>
> **Student 2**: The blob represents the electron itself, since an electron is described by a wave packet that will spread out over time. The electron acts as a wave and will go through both slits and interfere with itself. That's why a distinct interference pattern will show up on the screen after shooting many electrons.
>
> **Student 3**: Quantum mechanics is only about predicting the outcomes of measurements, so we really can't know anything about what the electron is doing between being emitted from the gun and being detected on the screen.

Ask students to read each statement one at a time, and respond before moving on to the next statement. Are student responses to the essay question consistent with their earlier descriptions of the experiment? Are student responses consistent with their earlier descriptions of atomic electrons? If no, why not? Is the student aware of inconsistencies?

QUESTIONS REGARDING INTERPRETATIONS OF QUANTUM MECHANICS
Is the student aware there are multiple interpretations of quantum mechanics? Can they name any of them or describe their features? Has the student heard of the Copenhagen Interpretation, and can they describe what it entails? Does the student know what the word *determinism* means within the context of physics? Did they have an opinion as to how they think their instructor would have wanted them to respond to earlier interview questions?

RESPOND TO THE STATEMENT IN AGREEMENT OR DISAGREEMENT AND EXPLAIN REASONING: *It is possible for physicists to carefully perform the same experiment and get two very different results that are both correct.*

RESPOND TO THE STATEMENT IN AGREEMENT OR DISAGREEMENT AND EXPLAIN REASONING: *The probabilistic nature of quantum mechanics is mostly due to the limitations of our measurement instruments.*

## A.2. INTERVIEW NARRATIVE AND DISCUSSION

Considering the contextual nature of student responses, we began each interview in the present study by asking students to simply describe an electron, in words or pictures, thereby allowing students to provide their own context. Three of the Quantum students differed from most other students in that their initial descriptions of electrons did not reference atoms – instead, these three students first described electrons as "wave packets", or "packets of energy". Fifteen of the remaining sixteen students described electrons first and foremost as constituents of atoms, and all but one of the remaining four students eventually discussed electrons within the context of atoms without any prompting from the interviewer. This approach allowed for an initial exploration of student beliefs about atomic electrons without the need for explicit questioning.

We found that a large majority of students used a planetary model as a first-pass description of atoms, though every student claimed to be aware that the Bohr model has limitations and is not entirely accurate; most students eventually said an atomic electron is more properly described by an *electron cloud*, or a *cloud of probability*. At this point, there were a number of opportunities to explore student ontologies: Do electrons exist as localized particles within that cloud of probability? Does the electron cloud represent something physical? Is it only a mathematical tool describing the probabilities for where an electron could be found? If the electron is acting like a wave, what is it that is waving? Is there anything physically oscillating in space? Once a clear picture had been established of how each student thought of electrons in the context of atoms, students were then asked to agree or disagree (and then explain their reasoning) with a statement from the online survey: *An electron in an atom has a definite but unknown position at each moment in time.* In every case, interview responses to this statement were consistent with the descriptions of electrons that students had just given.

Students were then asked to describe the setup for the double-slit experiment and what is observed. Every student discussed the observed fringe pattern in terms of the constructive and destructive interference of waves, and all but one student (R2) knew that attempts to determine which-path information would disrupt the interference pattern. All nineteen students were also aware the experiment could be run with single quanta, and that an interference pattern would still develop over time. These initial questions were meant to establish that each student had sufficient content knowledge regarding the double-slit experiment, so that a meaningful discussion concerning the implications of the results would be possible. From these responses, we conclude that no link could be discerned between a student's understanding of the particulars of the double-slit experiment, and their preferred interpretation of its results.

At this point in the interview, there were again opportunities to explore student stances on our themes of interpretation without the need for explicit questioning. Regardless of their preferred interpretation, every student explained the fringe pattern in terms of interference. If they thought of the experiment in terms of localized particles, what did they feel was the source of interference? If they preferred a wave description of quanta, how did students explain why single quanta are detected at localized points? Is it possible for a particle to pass through both slits simultaneously? Does a wave-packet description of an individual particle reflect an ignorance of the true state of that particle? As with the discussion of atomic electrons, by this point a clear picture had been established regarding how each student would interpret the double-slit experiment. Students were then asked to read and respond to statements made by three fictional students concerning their interpretations of the double-slit experiment. (See Appendix A) In the interviews, students were asked to read each statement, one at a time, and respond before moving on to the next statement.

Here, students did not necessarily respond to the essay question on the double-slit experiment in a manner consistent with their view on atomic electrons: some switched between Realist and Quantum perspectives, though no student applied first a Quantum perspective on atomic electrons and then a Realist perspective on the double-slit experiment. Nor did students necessarily remain consistent *within* their responses to the individual statements from the essay question. Any apparent inconsistencies in student responses were then explored with further questioning.

The interviews then progressed to a more explicit discussion of interpretation, with the intent of gauging student awareness of interpretive themes as actual themes of interpretation (as opposed to scientific fact). Had students heard of the Copenhagen Interpretation, and could they describe what it entails? Could students name any of the various interpretations of quantum mechanics, or describe their features? Did students know what the word *determinism* means within the context of physics? For all but a handful of students, the answers to the above questions were repeatedly: No. Most every student claimed to be aware that there are multiple interpretations of quantum mechanics, but very few could name even one of them. This pattern of student responses is consistent with our classroom observations: the instructors for the students interviewed rarely (if ever) discussed interpretations of quantum phenomena as actual interpretations; nor did these instructors devote a significant amount of class time (if any) to a discussion of the Copenhagen Interpretation, or any other interpretation of quantum mechanics.

To conclude the interviews, students were asked to respond in agreement or disagreement (and to provide their reasoning) to the following two statements:

- It is possible for physicists to carefully perform the same experiment and get two very different results that are both correct.
- The probabilistic nature of quantum mechanics is mostly due to the limitations of our measurement instruments.

Our intent was to explore whether and how students distinguish between the experimental uncertainties of classical physics and the fundamental uncertainties of quantum measurements, with the goal of gauging student perspectives on questions of determinism and probability. Surprisingly, all but one student (QR4) discounted the notion that a more complete (i.e non-statistical) description of quantum measurements would be possible. Student QR4 expressed a belief that scientists would eventually be able to predict the outcomes of individual quantum measurements with ever increasing accuracy. The remaining eighteen students either stated that the behavior of quantum particles is inherently probabilistic, and/or invoked the Uncertainty Principle as placing a fundamental limit on what can be known about quantum systems. Although realist/statistical perspectives have been historically associated with underlying notions of determinism, we find that practically none of our nineteen interview participants favored such ideas in the context of quantum physics. This finding is consistent with prior research showing that, not only do the perspectives of classical physics students become more deterministic over time, the perspectives of these same students become more probabilistic following a modern physics course. [1] For these reasons, a characterization of individual student perspectives with respect to determinacy could only be inferred from their stances on the other issues, and so is not included in this study.